\documentstyle[aps,twocolumn]{revtex}
\begin{document}
\voffset 0.5in
\draft
\wideabs{
\title{Relativistic Fluctuations and Anomalous Darwin Terms in Superconductors}
\author{Klaus Capelle}
\date{\today}
\address{Instituto de Qu\'{\i}mica de S\~ao Carlos, Departamento de 
Qu\'{\i}mica e F\'{\i}sica Molecular, Universidade de S\~ao Paulo,\\
Caixa Postal 369, S\~ao Carlos, 13560-970 SP, Brazil\\
and\\
Institut f\"ur Theoretische Physik, Universit\"at W\"urzburg,
Am Hubland, D-97074 W\"urzburg, Germany
}
\maketitle
\begin{abstract}
The anomalous Darwin term, one of the recently derived relativistic 
corrections to the conventional theory of superconductivity, is analysed 
in detail.  A new derivation of this term, much simpler than that
given originally, is presented and used to interpret the term physically,
in terms of fluctuations of the relativistic wave packet describing the paired 
electrons.
Several puzzling features of the original derivation find a simple explanation
in terms of this reinterpretation. 
The question of observability of the anomalous Darwin term, and its relation
to the conventional Darwin term and to earlier 
proposed superconducting Darwin terms are also discussed.
\end{abstract}
}
\newcommand{\be}{\begin{equation}}
\newcommand{\ee}{\end{equation}}
\newcommand{\bea}{\begin{eqnarray}}
\newcommand{\eea}{\end{eqnarray}}
\newcommand{\bi}{\bibitem}

\newcommand{\ep}{\epsilon}
\newcommand{\s}{\sigma}
\newcommand{\p}{{\bf \pi}}
\newcommand{\D}{\Delta}
\newcommand{\r}{({\bf r})}
\newcommand{\rp}{({\bf r'})}
\newcommand{\rrp}{({\bf r},{\bf r'})}
\newcommand{\xR}{({\bf x},{\bf R})}

\newcommand{\ua}{\uparrow}
\newcommand{\da}{\downarrow}
\newcommand{\la}{\langle}
\newcommand{\ra}{\rangle}
\newcommand{\dg}{\dagger}

\newcommand{\lmt}{\left(\begin{array}{cc}}
\newcommand{\lmf}{\left(\begin{array}{cccc}}
\newcommand{\rmat}{\end{array}\right)}
\newcommand{\lvec}{\left(\begin{array}{c}}
\newcommand{\rvec}{\end{array}\right)}

Recently it was demonstrated that a number of interesting and potentially 
observable relativistic effects exist in superconductors. The corresponding
relativistic theory of superconductivity was proposed in
Ref. \cite{physlett}, developed in detail in Refs. \cite{prb1} and 
\cite{prb2}, and briefly summarized in Ref. \cite{physica}.
A recent partial review is found in the textbook Ref. \cite{strange}. 
Among the predictions of the theory are the existence of new types of
unconventional superconducting order parameters (not derivable from 
Schr\"odinger's equation but only from the Dirac equation), and weakly 
relativistic correction terms to the conventional theory of superconductivity.
The present report is concerned with the physics of one of these correction 
terms.

Two essentially different types of such terms turned out to exist.
The first class are those corrections known from the application of 
relativity to other areas of solid-state or atomic physic, such as the
conventional kinetic-energy, spin-orbit, and Darwin corrections.
The latter two terms depend crucially on the presence of an external
electrostatic potential (in the present case a crystal lattice), and are zero 
for free particles.
The second class comprises terms (of the same order in $1/c$ and formally of a 
similar structure as those in the first class), which contain the 
superconducting pair potential in place of the lattice potential.
These terms depend crucially on the presence of a pair potential, and are 
zero for unpaired particles.
The superconducting counterparts to the spin-orbit and Darwin terms were
termed the anomalous spin-orbit and Darwin terms, respectively
\cite{physlett,prb1,prb2}.

Having derived these terms, one is now confronted with the task to 
analyze their consequences and assess their observability.
As a first step of this program the conventional and the
anomalous spin-orbit coupling have recently been investigated in detail 
in the framework of the theory of dichroism in superconductors
\cite{dichro1,dichro2}. It turned out that experimental results can be 
qualitatively understood on the basis of that theory \cite{dichro2}.
Another recent application of the theory is the proof by Strange 
\cite{strange} that the London equation governing the electrodynamics of simple
superconductors (which is usually derived in a nonrelativistic way), can be 
derived directly from the relativistic theory of superconductivity.

The purpose of the present report is to continue with this program by
taking a look at another term of the second class, namely the anomalous 
Darwin term.
A very simple rederivation of that term is given, which does not rely on
the formalism of Refs. \cite{prb1} and \cite{prb2}.
This rederivation provides a, previously unavailable, 
physical interpretation of that term, which is used to explain several 
puzzling features of the original derivation.
Conceptual consequences and the question of observability are discussed, as is
the relation of the anomalous Darwin term to similar terms, proposed by 
other authors.

The starting point is the standard Bogolubov-de Gennes equation 
\cite{dege,ketterson}
of the microscopic theory of superconductivity, which for spin-dependent 
potentials becomes a $4\times4$
matrix equation. For ease of notation this equation is here condensed in 
$2\times2$ form, by introducing the Pauli matrix $\hat{\sigma}_y$ and the 
unit matrix $\hat{I}$. For singlet superconductors in the absence of magnetic 
fields the equation then takes the form
\bea
\lmt \hat{h}\r \hat{I} & i\hat{\sigma}_y \D\rrp \\
-i\hat{\sigma}_y \D^*\rrp & -\hat{h}\r \hat{I} \rmat
\lvec u_{n\s}\r \\ v_{n\s}\r \rvec
\nonumber \\
=E_{n\s} \lvec u_{n\s}\r \\ v_{n\s}\r \rvec,
\label{bdg}
\eea
where $\hat{h}\r = -\hbar^2\nabla^2/(2m) + v\r$ is the single-particle
Hamiltonian describing the normal state, and $v\r$ is the lattice potential, 
including a mean field \cite{dege} or Kohn-Sham \cite{ogk1} term arising 
from the Coulomb interaction.
$\D\rrp$ is the (generally nonlocal) pair potential of the superconductor,
and $u_{n\s}\r$ and $v_{n\s}\r$ are the particle and hole amplitudes,
coupled by this potential.
This equation, which is usually derived in a nonrelativistic way 
\cite{dege,ketterson}, can be identified as
an approximation to lowest order in $v/c$ 
(where $v$ is a typical particle velocity and $c$ the velocity of light) to the 
relativistic Dirac-Bogolubov-de Gennes equation \cite{physlett,prb1}, which 
describes pairing on the basis of the Dirac equation.

To second order one finds various weakly relativistic correction terms
to Eq. (\ref{bdg}) \cite{physlett,prb2}.
The one of interest in the present work is the
so called anomalous Darwin term, which is to be added to the offdiagonal
entries of the matrix in Eq. (\ref{bdg}). This term is given by
\be
d_D({\bf x},{\bf R}) = i \hat{\sigma}_y\frac{\hbar^2}{8m^2c^2} 
\nabla_R^2 \Delta({\bf x},{\bf R}),
\ee
where ${\bf x}={\bf r}-{\bf r}'$ and ${\bf R}=({\bf r}+{\bf r}')/2$ 
are relative and center-of-mass coordinates
of the two electrons in the Cooper pair, respectively. For comparison 
purposes we also record the conventional Darwin term 
\be
v_D\r = \hat{I} \frac{\hbar^2}{8m^2c^2} \nabla_r^2 v\r,
\ee
which appears as a correction to the diagonal entries of Eq. (\ref{bdg}).
This term is usually derived by applying the Foldy-Wouthuysen
transformation or the Pauli approximation to the Dirac equation
\cite{strange,rose,sakurai}. 
In the same way, the anomalous Darwin term 
(together with the other anomalous corrections) follows from the 
Dirac-Bogolubov-de Gennes equation \cite{physlett,prb1,prb2}. 

In the conventional case a much simpler alternative derivation exists, which 
is based on the fact that a wave packet formed from solutions to the
Dirac equation performs a very rapid oscillatory motion
with amplitude $\delta{\bf r} \approx \lambda_c:=\hbar/(mc)$,
where $\lambda_c$ is the reduced Compton wavelength. 
This so called 'Zitterbewegung' (trembling motion) 
results from the interference of the positive-energy and negative-energy 
components of the wave packet \cite{strange,rose,sakurai,itzzub}.
On the average, the particle thus feels the potential spread out over 
a region of order $\delta{\bf r}$, instead of its value at
${\bf r}$. The resulting average change in energy is easily estimated to be
\cite{rose,sakurai,itzzub}
\bea
\la \delta v\r \ra = \la v({\bf r} + \delta{\bf r}) - v\r \ra
\nonumber \\
=\left\la \sum_i \delta r_i \frac{\partial V}{\partial r_i} +
\frac{1}{2} \sum_{ij} \delta r_i \delta r_j 
\frac{\partial^2 V}{\partial r_i\partial r_j} + \ldots\right\ra
\nonumber \\
\approx  
\frac{1}{6} (\delta{\bf r})^2 \nabla_r^2 v\r
\approx \frac{\hbar^2}{6m^2c^2} \nabla_r^2 v\r,
\label{convest}
\eea
where it was assumed that the fluctuations do not have a preferred 
direction in space. The resulting average potential $\la \delta v\r \ra$
is, up to a numerical factor $6/8$, that of the conventional 
Darwin term $v_D\r$. (The more accurate estimate 
$(\delta{\bf r)}^2 \approx (3/4) \lambda^2_c$, occasionally found in the 
literature \cite{rose}, reproduces the Darwin term exactly.)
The above line of reasoning is very powerful, in spite of its simplicity.
It requires, for example, only a slight generalization to yield an
accurate estimate for the Lamb shift (see, e.g., Ref. \cite{itzzub}, p. 80),
which is a much more subtle relativistic effect than the Darwin term.

We now proceed by generalizing the same procedure to the superconducting
case.
Due to the presence of the pair potential $\D\rrp$ the particles now
fluctuate in two potentials, each of which affects the single-particle
energy. In a first approximation the fluctuations in the potentials
$v\r$ and $\D\rrp$ can be treated as independent. It thus remains to calculate
the average change of energy due to the fluctuations in the pair 
potential.
In the same way as above one finds for the fluctuations of the center-of-mass
coordinate of the pair
\bea
\la \delta \D\xR \ra = \la \D({\bf x},{\bf R} + \delta{\bf R}) - \D\xR \ra =
\nonumber \\
\left\la \sum_i \delta R_i \frac{\partial \D\xR}{\partial R_i} +
\frac{1}{2} \sum_{ij} \delta R_i \delta R_j
\frac{\partial^2 \D\xR}{\partial R_i\partial R_j} + \ldots\right\ra
\nonumber \\
\approx 
\frac{1}{6} (\delta{\bf R})^2 \nabla_R^2 \D\xR.
\label{darwinest}
\eea
Using again the estimate $(\delta{\bf R})^2 \approx \lambda^2_c$ 
for the amplitude of the Zitterbewegung, we find the effective potential
\be
\la \delta \D\xR \ra = \frac{\hbar^2}{6m^2c^2}\nabla_R^2 \D\xR,
\label{adt}
\ee
which is, up to the same numerical prefactor as in the conventional case, 
the anomalous Darwin term, as obtained from the microscopic theory 
\cite{physlett,prb2}.
This simple rederivation now allows us to draw a number of conclusions 
regarding the nature of the anomalous Darwin term.

First, note that the anomalous Darwin term is one of the main
results of Refs. \cite{physlett} and \cite{prb2}, obtained
there after lengthy algebraic manipulations based on the general theory
of Ref. \cite{prb1}. The fact that one can find a much simpler 
rederivation of this term, not based on the general equations of those
papers, gives additional confidence in that theory.

Second, from the point of view of those references it remained a puzzle
why no anomalous Darwin term containing derivatives with respect to the 
relative coordinate exists, in particular since
there is such a term in the case of the anomalous spin-orbit coupling
(there are, effectively, two anomalous spin-orbit terms, one containing
$\nabla_x \D\xR$, the other $\nabla_R \D\xR$ \cite{prb2}).
It was already verified in Ref. \cite{prb2} that the hypothetical
existence of an anomalous Darwin term containing $\nabla_x^2 \D\xR$
is incompatible with the correct local limit, but the 
physical reason for the absence of this term was not apparent from the
microscopic calculation of that work. 

The present rederivation of this term from relativistic
fluctuations of the paired particles in the pair potential suggests a
simple solution to this puzzle:
Once the electrons are paired, their fluctuations are not independent
anymore. The electrons in the
Cooper pair are locked into mutually time-conjugate single-particle states,
so that relative fluctuations within the pair are strongly suppressed.
We are thus dealing only with fluctuations of the center-of-mass coordinate
${\bf R}$ of the pair, but not of the relative coordinate ${\bf x}$.
Essentially, the entity fluctuating in the
{\em pair} potential is the pair as a whole.

Additional insight into the nature of the anomalous Darwin term is gained
by comparing it with
the superconducting Darwin terms derived in Refs. \cite{nrdarwin1}
and \cite{nrdarwin2}. In these papers it is pointed out that even in a
nonrelativistic situation the Bogolubov-de Gennes equations do have
negative-energy solutions (representing the condensate), separated by an 
energy gap from the positive-energy states (the excited quasiparticles,
representing broken pairs).
By approximately decoupling these two types of states via a
Foldy-Wouthuysen transformation, purely
superconducting counterparts to the Zitterbewegung and the conventional
Darwin term were found. These superconducting Darwin terms are superficially
similar to the one discussed here.
However, although of superconducting origin, they do not contain the
pair potential, but the ordinary lattice potential, and are thus closer
related to the conventional Darwin term than to the anomalous one.
Furthermore, not being of relativistic origin, they do not carry the
relativistic prefactor $\hbar^2/(8m^2c^2)$ of the Darwin terms discussed
here.
Finally, the superconducting Darwin terms of Refs. \cite{nrdarwin1}
and \cite{nrdarwin2}
do not appear in the standard Bogolubov-de Gennes equation (\ref{bdg}),
but only in an approximate form of that equation, in which
the excited quasiparticles were approximately decoupled from the
condensed pairs \cite{nrdarwin1,nrdarwin2}.
This is completely analogous to the case of the Dirac
equation, from which the conventional and the anomalous Darwin terms are 
derived by approximately decoupling electrons from positrons \cite{footnote1}.

All of the above considerations were based on the relativistic generalization
of the BCS (singlet) order parameter. However, the general theory of 
Refs. \cite{prb1} and \cite{prb2} also provided the form of the relativistic
corrections to triplet superconductors, and predicted the possibility of a new
type of order parameter, involving charge conjugation instead of time reversal
as pairing symmetry.
Recently these terms have been analysed in more detail \cite{miguel}.
While for triplet superconductors one finds similar anomalous Darwin terms as
in the singlet case, for the order parameters involving one charge conjugation
operation no anomalous Darwin terms are found at all, not even 
one involving the center-of-mass coordinate. 

The above rederivation of the anomalous Darwin term provides a simple 
explanation for this absence.
It was already pointed out in Ref. \cite{prb2} that pairs described by an 
order parameter involving one charge conjugation operation are 
intrinsically relativistic; the corresponding pair potential vanishes
in the nonrelativistic limit. 
On the other hand, the only place in which relativity enters Eqs. 
(\ref{convest}) and (\ref{darwinest}) is as the origin of the fluctuations, 
$\delta{\bf r}$ and $\delta{\bf R}$. In particular, to order $1/c^2$ or,
equivalently, $(\delta{\bf R})^2$, the pair potential in Eqs. (\ref{darwinest}) 
and (\ref{adt}) is the nonrelativistic pair potential.
Consequently, for order parameters involving one charge conjugation
the pair effectively fluctuates, to order $(\delta{\bf R})^2$, only in the 
crystal potential, but not in the pair potential. 
Hence there are no associated anomalous Darwin terms for these order 
parameters. This is the result also found after considerable algebraic 
manipulations from the microscopic theory \cite{miguel}.

We now return to the conventional singlet case and proceed with a tentative 
discussion of 
the possibility of experimental observation of the anomalous Darwin term.
The conventional Darwin term is normally a rather small effect.
This is not only due to the prefactor $1/c^2$, but also related to the
fact that, unlike the spin-orbit term, which breaks rotational symmetry, the 
conventional Darwin term does not break any nonrelativistic symmetries.
In the case of the anomalous Darwin term the situation is similar.
Although
this term can break symmetries because it contains the pair potential, which 
may break the symmetry of the crystal lattice in so called unconventional 
superconductors \cite{sigue}, this symmetry breaking is of entirely
nonrelativistic origin and manifests itself not only in the relativistic
correction terms, but already in the nonrelativistic theory based on
Eq. (\ref{bdg}).
It is thus expected that experimental observation of the anomalous Darwin
term is much harder than that of the anomalous spin-orbit coupling term
(which may be detected, e.g., by magnetooptical means \cite{dichro1,dichro2}).

An essential difference between the anomalous and the conventional Darwin 
term, which may facilitate experimental detection of the former, is 
the temperature dependence of the pair potential, which is absent from the
conventional Darwin term, but intrinsic to the anomalous one.
Another important difference is related to the derivatives of the potentials. 
While the expectation value of the conventional ionic Darwin term, originating
from the ions at positions ${\bf R}_i$, is nonzero 
only for s-states, which are nonvanishing at the position of the nuclei
\cite{strange,sakurai},
\be
\nabla_r^2v_{ion}\r = \sum_i \nabla_r^2 \frac{1}{|{\bf r} - {\bf R}_i|}
=-4\pi \sum_i \delta^{(3)}({\bf r} - {\bf R}_i),
\label{clb}
\ee
the anomalous Darwin term can also influence other states, since the pair 
potential $\D\xR$ is not a simple sum of Coulomb-like contributions.
Any signature of a Darwin term for non s-states (e.g., the p-state 
contribution to the conduction band in simple metals, or the partially 
filled d-shells of transition metals) with the characteristic
temperature dependence of the pair potential, would constitute an unequivocal 
identification of the anomalous Darwin term. Such an identification may come 
from high-resolution photoemission or tunneling experiments, detecting
the small temperature dependent energy shifts produced by the anomalous Darwin 
term. 

A direct comparison of the order of magnitude of the pair potential $\Delta$
(of the order of $0.1$ meV in BCS superconductors and about hundred times
larger in high-temperature superconductors) with $mc^2$ ($511$ keV) is
rather discouraging, but the fact that the anomalous Darwin
term affects all states involved in pairing (and not only the s-states)
partially compensates for this difference in energies. 
Furthermore, in solid-state situations the relativistic prefactor 
$1/c^2$ does not always imply smallness of the effects, because the large
number of involved particles can make up for this factor
\cite{footnote2}, in particular in superconductors, where the particles 
act coherently. The contribution of the anomalous spin-orbit coupling (which 
is of the same order in $1/c$ as the anomalous Darwin term) to the 
magnetooptical response of a superconductor is, for example, hugely
enhanced near the critical temperature by the superconducting coherence
factors \cite{dichro1,dichro2}.

Due to the second derivative of the pair potential with respect to the
center-of-mass coordinate of the Cooper pair, the anomalous Darwin term is 
larger in situations in which the pair potential varies rapidly in space.
Examples are thin films, surfaces, or vortices, in particular in systems
with a short coherence length. An ideal system for observations would thus
be a copper-oxide superconductor, since these simultaneously have a very short 
coherence lenght ($\xi$ is only about $10^3$ Compton wave lengths!), a large 
pair potential (evidenced by the large energy gap), and very heavy elements in 
the lattice (which generally favours relativistic effects).

As long as such experiments have not been performed, the main results of
the present paper are of a conceptual nature:
(i) The much simpler rederivation of one of the main
results of Refs. \cite{prb1} and \cite{prb2} gives additional 
confidence in the more complex formalism of those references.
(ii) The interpretation in terms of rapid spatial fluctuations of the Cooper 
pair in the pair potential elucidates the physical origin of the 
anomalous Darwin term, which was not apparent from the earlier work.
It also yields a simple explanation for the previously puzzling fact 
that the anomalous Darwin term contains only derivatives with respect to the 
center-of-mass coordinate of the Cooper pair, and not also derivatives with 
respect to the relative coordinate.
(iii) The same interpretation also explains the
absence of any type of anomalous Darwin terms for relativistic order
parameters involving one charge conjugation.
(iv) The temperature dependence of the anomalous Darwin term, together with
its effect on other than s-states, were argued to provide a clear signature 
of this term.
Thin films of high-temperature superconductors (or other superconductors with 
a short coherence length and a large pair potential) were suggested as ideal 
systems to look for observable consequences of the anomalous Darwin term. 

{\bf Acknowledgments}
Valuable discussions with E. K. U. Gross, L. N. Oliveira, M. Marques, 
and B. L. Gy\"orffy are gratefully acknowledged.
Part of this work was supported by the German Science Foundation (DFG), part
by the Science Foundation of the State of S\~ao Paulo, Brazil (FAPESP).

\end{document}